# From Affine $A_4$ to Affine $H_2$: Group Theoretical Analysis of Five-fold Symmetric Tilings

**Nazife Ozdes Koca**[a,*], **Ramazan Koc**[b], **Mehmet Koca**[c] **and Rehab Al-Reasi**[a]

[a]Department of Physics, College of Science, Sultan Qaboos University, P.O. Box 36, Al-Khoud 123, Muscat, Sultanate of Oman, *Correspondence e-mail: nazife@squ.edu.om
[b]Department of Physics, Gaziantep University, Gaziantep, Turkey
[c]Department of Physics, Cukurova University, Adana, Turkey, retired

## ABSTRACT

The projections of the lattices, may be used as models of quasicrystals, and the particular affine extension of the $H_2$ symmetry as a subgroup of $A_4$, discussed in the work, presents a different perspective to 5-fold symmetric quasicrystallography. Affine $H_2$ is obtained as the subgroup of the affine $A_4$. The infinite discrete group with local dihedral symmetry of order 10 operates on the Coxeter plane of the root and weight lattices of $A_4$ whose Voronoi cells tessellate the 4D Euclidean space possessing the affine $A_4$ symmetry. Facets of the Voronoi cells of the root and weight lattices are identified. Four adjacent rhombohedral facets of the Voronoi cell $V(0)$ of $A_4$ project into the decagonal orbit of $H_2$ as thick and thin rhombuses where long diagonals of the rhombohedra serve as reflection line segments of the reflection operators of $H_2$. It is shown that the thick and thin rhombuses constitute the finite-fragments of the tiles of the Coxeter plane with the action of the affine $H_2$ symmetry. Projection of the Voronoi cell of the weight lattice onto the Coxeter plane tessellates the plane with four different tiles: thick and thin rhombuses with different edge lengths obtained from the projection of the square faces and two types of hexagons obtained from the projection of the hexagonal faces of the Voronoi cell. Structure of the local dihedral symmetry $H_2$ fixing a particular point on the Coxeter plane is determined

**Keywords**: Affine groups, quasicrystals, aperiodic tilings, lattices, projections of polytopes

## 1. Introduction

Voronoi cells of the root and weight lattices of the Coxeter-Weyl groups [Conway & Sloane, 1988, 1991] of rank N tessellate the N-dimensional Euclidean space facet to facet. It is then important to determine the facets of the Voronoi cells [Koca et. al., 2012, Koca et. al., 2014, Koca et. al., 2018]. Equally important is the structure of the affine Coxeter-Weyl group [Humphreys, 1992], which constitutes the symmetry of the associated lattices as well as their tessellations by their Voronoi cells. Many of these groups admit non-crystallographic groups as finite subgroups, which could be useful in the classification of quasicrystals [de Bruijn, 1981; Whittaker & Whittaker, 1987; Koca et.al. 2015]. The well-known non-crystallographic groups $H_2$, $H_3$ and $H_4$ are the respective subgroups of the Coxeter groups $A_4$, $D_6$ and $E_8$. The affine extensions of the groups $H_2$, $H_3$ and $H_4$ similar to the affine extensions of the Coxeter-Weyl groups were first discussed in the reference [Patera & Twarock, 2002] and possible other affine extensions have been discussed in the reference [Dechant et.al., 2012]. Most popular of these affine groups is the affine $A_4$ admitting $H_2$ as a point symmetry, the dihedral group of order 10 [Baake et al, 1990]; see also [Baake & Grimm, 2013] for further discussions. The Coxeter group $D_6$ embeds the icosahedral group $H_3$ as a subgroup [Kramer, 1993] and the exceptional Coxeter group $E_8$ embeds the non-crystallographic Coxeter group $H_4$ as a subgroup [Coxeter, 1973; Koca et al, 2001]. It seems the affine extension of $H_3$ may play important roles in quasicrystals and in viral structures [Keef & Twarock, 2008; Indelicato et. al., 2012; Dechant et al, 2013; Salthouse et. al., 2015; Zappa et. al., 2016]. Structural transitions in quasicrystals induced by transitions in higher dimensional lattices are discussed in [Giulliana et.al., 2012]. Earlier, by employing the cut and project technique, we have projected the quaternionic root lattice of $A_4$ [Koca et. al., 2014] tessellated by the Delone cells on to the Coxeter plane. In what follows we determine the affine $H_2$ as a subgroup of affine $A_4$ and apply it on the projections of the Voronoi tessellations of the root lattice $A_4$ and the weight lattice $A_4^*$. We extend the graph-folding technique of Shcherbak [Shcherbak, 1988] to the graph-folding of affine $A_4$ to determine the affine structure of $H_2$ as detailed in Appendix A. For further references on graph-folding see the reference [Dechant, P. P., 2017). Facets of the Voronoi cells of the root and weight lattices after projection onto the Coxeter plane turn out to be important as they involve the affine reflection line segments of the group $H_2$. This technique allows the explicit construction of the local structures of the $H_2$ group connected by the translation element leading to the five-fold tessellations of the Coxeter plane. By employing this technique, we obtain two types of tessellations: one with the thin and thick rhombus tilings projected from the Voronoi tessellations of the root lattice and the other is obtained from the Voronoi tessellations of the weight lattice admitting four different tiles, thin-thick rhombuses with different edge lengths and two hexagons with two edge lengths in proportion to golden ratio. The latter tessellation has not been discussed earlier in elsewhere.

The paper is organized as follows. In section 2 we introduce the affine extension of the group $A_4$, especially taking into account its action on the rhombohedral facets of the Voronoi cell of the root lattice. Section 3 deals with the study of affine subgroup $H_2$ and its action on the projection of the Voronoi cell onto the Coxeter plane. We point out that there is a unique tiling of the projected decagon with thin and thick rhombuses in which two points are left invariant under the local $H_2$ symmetry. Around these fixed points, we obtain two different centrally symmetric Penrose tilings by applying affine $H_2$ subgroup. Section 4 is devoted to the discussion of the structure of the Voronoi cell of the weight lattice $A_4^*$, which is also called the permutohedron of order 5 admitting truncated octahedra and hexagonal prisms as 3D facets. We project the permutohedron onto the Coxeter plane and tessellate the plane by thin-thick rhombuses and thin-thick hexagons by applying the affine $H_2$. In section 5 we discuss our results and contrast with the existing five-fold symmetric tessellations of the plane. Appendix A discusses the explicit construction of the generators of the affine $A_4$ both in the formal structure and in an explicit $5 \times 5$ matrix representation leading to the $3 \times 3$ matrix representation of the generators of affine subgroup $H_2$.

## 2. Affine $A_4$ and projection of the Voronoi cell of the root lattice

The extended Coxeter-Dynkin diagram representing the affine extension of $A_4$ is shown in Fig. 1 where the roots are labeled by $\alpha_i = k_i - k_{i+1}$ (i=1, 2, 3, 4) and $\alpha_0 = -(\alpha_1 + \alpha_2 + \alpha_3 + \alpha_4) = k_5 - k_1$ and the norm of the roots is $\sqrt{2}$.

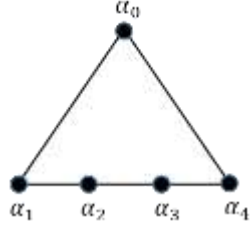

Figure 1. Extended Coxeter-Dynkin diagram of $A_4$.

Here the vectors $k_i$ represent the vertices of the 4-simplex (5-cell) and satisfy the relations [Koca et al, 2019],

$$k_1 + k_2 + k_3 + k_4 + k_5 = 0, \; (k_i, k_i) = \frac{4}{5}, \; (k_i, k_j) = -\frac{1}{5}, (i \neq j). \tag{1}$$

The weight vectors $\omega_i$ of $A_4$ are obtained from the relation $\omega_i = \sum_{j=1}^{4}(C^{-1})_{ij}\alpha_j$ where $C_{ij} = (\alpha_i, \alpha_j)$ is the Cartan matrix and they can be expressed in terms of the vectors $k_i$ as

$$\omega_1 = k_1, \omega_2 = k_1 + k_2, \omega_3 = k_1 + k_2 + k_3 = -(k_4 + k_5), \omega_4 = -k_5. \tag{2}$$

The generator $r_i \coloneqq r_{\alpha_i}$ of $A_4$ reflects the vectors with respect to the hyperplane passing through the origin and orthogonal to the root $\alpha_i$ which acts on the vector $k_i$ as $r_i(k_i) = k_{i+1}$.

The vertices of the Voronoi cell $V(0)$, centered around the origin, of the root lattice is the union of the orbits of the weight vectors [Koca et al, 2019] which are given as the sets of vectors $\pm k_i$, $\pm(k_i + k_j), (i \neq j)$. To set the scene we will introduce the affine Coxeter-Weyl group in its general context [Humphreys, 1992], which includes the translation.

For a vector $\lambda \in V$ in the Euclidean space let $t(\lambda)$ represent the translation $t(\lambda)\mu = \mu + \lambda$. For any element $g \in GL(V)$, $gt(\lambda)g^{-1} = t(g\lambda)$ that can be obtained from the generalization of the reflection generator, which reflects the vectors with respect to the hyperplanes not necessarily passing through the origin. For each root $\alpha$ and for each integer $n$ the affine hyperplane is defined as the set of vectors $\lambda$

$$H_{\alpha,n} := \{\lambda \in V | (\lambda, \alpha) = n\}, \tag{3}$$

For $n = 0$, $H_{\alpha,0} = H_\alpha$ is the hyperplane through the origin orthogonal to the root $\alpha$. It is then clear that $H_{\alpha,n}$ can be obtained by translating $H_\alpha$ by $\frac{n}{2}\alpha$. The affine reflection with respect to the hyperplane $H_{\alpha,n}$ is defined as follows

$$r_{\alpha,n}(\lambda) := \lambda - ((\lambda, \alpha) - n)\alpha, \quad n \in \mathbb{Z}. \tag{4}$$

Note that for $n = 0$, $r_{\alpha,0} = r_\alpha$ and $r_{\alpha,n}$ fixes $H_{\alpha,n}$ pointwise and translate the origin to $n\alpha$ which can also be written as $r_{\alpha,n}(\lambda) = t(n\alpha)r_\alpha$. A similarity transformation by translation of a general reflection operator will read

$$t(\lambda)\, r_{\alpha,n} t(-\lambda) = r_{\alpha, n+(\lambda,\alpha)}. \tag{5}$$

For practical reasons, we will use the notation $(g, \lambda)$ for a general group element transforming an arbitrary vector $\mu \in V$ as $(g, \lambda)\mu = g\mu + \lambda$, where $g$ is the group element fixing the origin and $\lambda$ is a translation. This will be useful when we denote the general group element in the matrix notation. In this notation the inverse of the group element will be given by $(g, \lambda)^{-1} = (g^{-1}, -g^{-1}\lambda)$ and the unit element is $(1, 0)$.

Since we will discuss the projection of the Voronoi cell $V(0)$ of the root lattice we should determine its 3D facets. This has been discussed earlier [Koca et. al., 2012, Koca et. al., 2018] but here we will construct it in terms of the vectors $k_i$. Note that the Voronoi cell is the dual of the root polytope whose vertices are the set of vectors $k_i - k_j$, $(i \neq j)$ and can be obtained from the highest weight vector $\omega_1 + \omega_4 = k_1 - k_5$. Vertices of the 3D facet of $V(0)$ orthogonal to this root can be obtained from the vertices of the irregular tetrahedron $(\omega_1, \omega_2, \omega_3, \omega_4)$, the "roof" of the fundamental simplex [Conway & Sloane, 1991]. The fundamental simplex with vertices $(0, \omega_1, \omega_2, \omega_3, \omega_4)$ is one of the 120 cells of $V(0)$. Each cell can be regarded as a 4D pyramid based on the roof of the fundamental simplex. Another interesting fact is that the Delone cells whose vertices are all possible combinations of the orbits of $\omega_1 + \omega_4$ and $\omega_2 + \omega_3$ centralize the vertices of the Voronoi cell $V(0)$. The 3D facets of the Voronoi cell $V(0)$ are the rhombohedra whose centers are the halves of the roots $\frac{1}{2}(k_i - k_j)$. A typical rhombohedron whose centre is half the root $\frac{1}{2}(k_1 - k_5)$ can be obtained by applying the subgroup $<r_2, r_3>$ on the vertices $(\omega_1, \omega_2, \omega_3, \omega_4)$ of the roof leading to the set of vertices

$$<r_2, r_3>(\omega_1, \omega_2, \omega_3, \omega_4) = \{k_1, k_1 + k_2, k_1 + k_3, k_1 + k_4, -(k_2 + k_5), -(k_3 + k_5),$$
$$-(k_4 + k_5), -k_5\}. \tag{6}$$

These 8 vertices constitute a rhombohedron with identical rhombic faces as shown in Fig. 2.

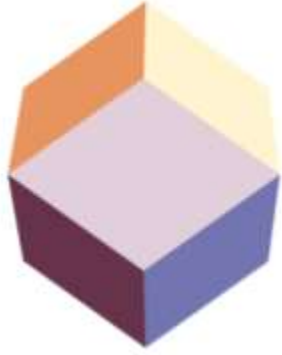

Figure 2. The rhombohedral facet of the Voronoi cell $V(0)$.

Note that any pair of the vectors $(k_i, k_j)$ forms a rhombus and any triple of vectors $(k_i, k_j, k_l)$ generates a rhombohedron. The Voronoi cell has 20 identical rhombohedral facets and the 20 4D pyramids with basis of rhombohedra constitute the Voronoi cell $V(0)$ and in turn each pyramid consists of 6 identical simplexes so the Voronoi cell $V(0)$ consists of 120 simplexes. Each rhombohedron of $V(0)$ is embedded in one of the hyperplane $H_{\alpha,1}$ where $\alpha$ is one of those 20 roots $(k_i - k_j)$. To give an example, the affine reflection fixing the rhombohedron piecewise with the center represented by $\frac{1}{2}\alpha_0 = \frac{1}{2}(k_5 - k_1)$ is given by the reflection generator $r_{\alpha_0,1}$. Therefore Fig. 1 represents the affine $A_4$ generated by the reflection generators $< r_1, r_2, r_3, r_4, r_{\alpha_0,1} >$. An arbitrary reflection element of the affine $A_4$ is then given by $r_{\alpha,n}$ as defined in (4).

First, we introduce the Coxeter plane before we discuss the details of the projection of the Voronoi cell $V(0)$. The Coxeter element of $A_4$ can be defined as the product of two reflection generators $R = R_1 R_2$, where $R_1 = r_1 r_3$ and $R_2 = r_2 r_4$. A better choice of the Coxeter element can be obtained from $R$ by a similarity transformation $R' = SRS^{-1} = r_1 r_2 r_3 r_4$ which permutes $k_i$ in the cyclic order where $S = r_4 r_3$. Then the simple roots would read $\alpha_i' = S\alpha_i$ and the corresponding reflection generators would be $r_i' = S r_i S^{-1}$. We can now represent the simple roots of the subgroup $H_2$ in two complementary subspaces $E_\parallel$ and $E_\perp$ as follows

$$E_\parallel: \beta_1 = \tfrac{1}{\sqrt{2+\tau}}(\alpha_1' + \tau\alpha_3'), \quad \beta_2 = \tfrac{1}{\sqrt{2+\tau}}(\tau\alpha_2' + \alpha_4'), \quad \tau = \tfrac{1+\sqrt{5}}{2},$$

$$E_\perp: \gamma_1 = \tfrac{-1}{\sqrt{2+\sigma}}(\alpha_1' + \sigma\alpha_3'), \quad \gamma_2 = \tfrac{-1}{\sqrt{2+\sigma}}(\sigma\alpha_2' + \alpha_4'), \quad \sigma = \tfrac{1-\sqrt{5}}{2}. \quad (7)$$

The Coxeter plane is taken as the plane $E_\parallel$. Defining the orthogonal unit vectors in 4D Euclidean space as

$$\tfrac{\beta_1 - \beta_2}{\sqrt{2(2+\tau)}} = \hat{x}, \quad \tfrac{\tau(\beta_1+\beta_2)}{\sqrt{2}} = \hat{y}, \quad \tfrac{\gamma_1-\gamma_2}{\sqrt{2(2+\sigma)}} = \hat{z}, \quad \tfrac{\sigma(\gamma_1+\gamma_2)}{\sqrt{2}} = \widehat{w}, \quad (8)$$

one obtains a representation of the vectors $k_j = \sqrt{\tfrac{2}{5}}(e^{i\frac{2\pi}{5}j}, e^{i\frac{4\pi}{5}j})$, $(j = 1,2,3,4,5)$ with a pair of complex components which are cyclically rotated by the Coxeter element $R' = r_1 r_2 r_3 r_4$. First and second complex components of $k_j$ represent the vector in $E_\parallel$ and $E_\perp$ respectively.

With this definition of the simple roots of $H_2$ given in two orthogonal spaces, the extended Coxeter graph in Fig. 1 can be decomposed as shown in Fig. 3.

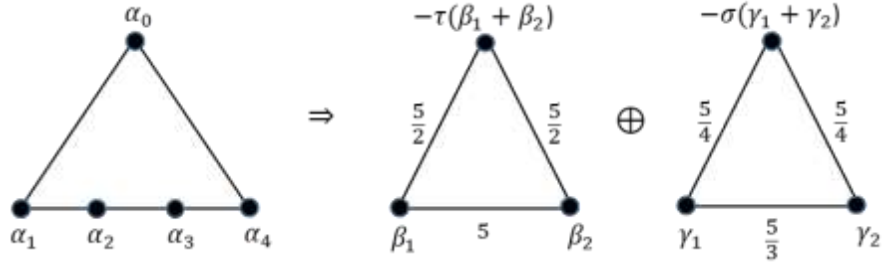

Figure 3. Decomposition of the extended Coxeter-Dynkin diagram of $A_4$ as two affine extensions of $H_2$ in two orthogonal spaces.

Note that the extended root of $A_4$ can be written as the linear combination of the extended roots of $H_2$

$$-(\alpha_1' + \alpha_2' + \alpha_3' + \alpha_4') = -\frac{\tau(\beta_1+\beta_2)}{\sqrt{2+\sigma}} - \frac{\sigma(\gamma_1+\gamma_2)}{\sqrt{2+\tau}} = k_4 - k_1. \qquad (9)$$

It is clear from (9) that the projection of the root $k_4 - k_1$ onto the Coxeter plane is the scaled copy of the root $-\tau(\beta_1+\beta_2)$ which will be discussed in Section 3. We shall project the Voronoi cells of the root and weight lattices onto the space $E_\parallel$. For this, we work with the root system generated by the simple roots $\beta_1$ and $\beta_2$ and discuss them in Section 3 in connection with the affine $H_2$. The automorphism group of the root system of $A_4$ is of order 240 and it is the semi-direct product of the symmetric group $S_5$ with the Dynkin-diagram symmetry $Z_2$. This leads to the automorphism group of the root system of $H_2$ in the form of semi-direct product $H_2: Z_2$, a group of order 20. The 30 vertices of the Voronoi cell $V(0)$ of the root lattice project onto the Coxeter plane as three concentric circular orbits of the group $H_2: Z_2$ with radii $\sqrt{\frac{2}{5}}\tau^{-1}, \sqrt{\frac{2}{5}}, \sqrt{\frac{2}{5}}\tau$, each containing 10 vectors which can be represented as the cyclic permutations of the sets of vectors $\pm(k_1 + k_3)$, $\pm k_1$ and $\pm(k_1 + k_2)$ respectively. We use the same notation $k_j$ for the projected component $k_j = \sqrt{\frac{2}{5}}e^{i\frac{2\pi}{5}j}$ to avoid the frequent use of the notation $k_{\parallel j}$. We display in Fig. 4 the only possible projection of the Voronoi cell $V(0)$ into a decagon up to a cyclic permutation of order 5.

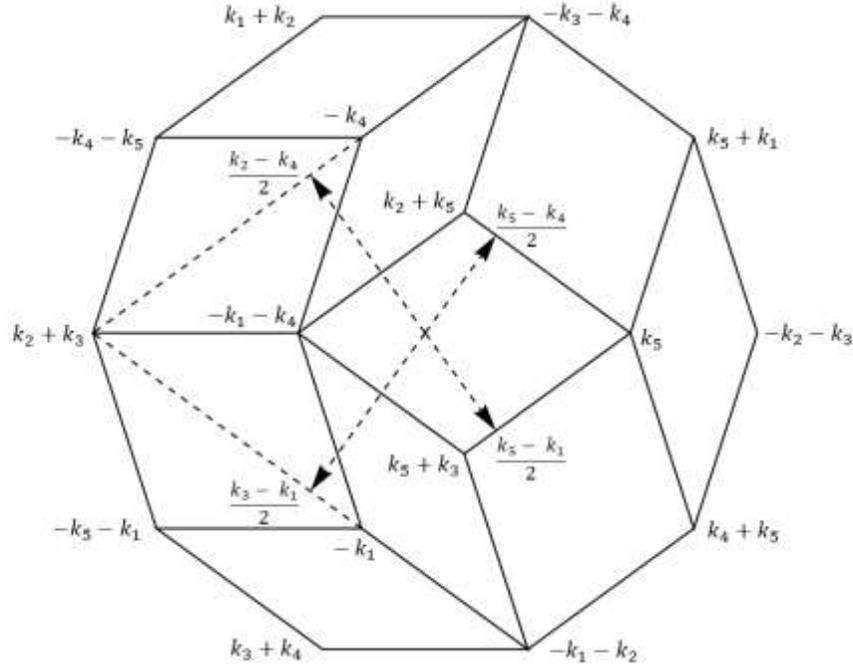

Figure 4. Projection of the Voronoi cell $V(0)$ of the root lattice $A_4$ onto the Coxeter plane.

From Fig. 4 one can easily see that thick and thin rhombuses are the faces of four rhombohedra which are pairwise adjacent to each other. The centers of those projected rhombohedra are pointed out as $\frac{k_5-k_4}{2}$, $\frac{k_2-k_4}{2}$, $\frac{k_3-k_1}{2}$, $\frac{k_5-k_1}{2}$. The line segments orthogonal to these vectors are the reflection symmetry axes of the affine $H_2$ group as we will discuss in the next section. Each of these four projected rhombohedra is symmetric with respect to the line segment bisecting the corresponding roots. Cyclic rotation of the decagon by the Coxeter element $R'$ will produce five copies of the same figure without changing the distribution of the rhombuses exhausting all possible quadrupole partitioning of the root system of $A_4$. Hence the decagon in Fig. 4 is the only projection of $V(0)$ onto the Coxeter plane. The other possible tilings of the decagon with equal number of thick-thin rhombuses cannot be interpreted as the projections of the rhombohedra.

## 3. Affine $H_2$ as a subgroup of affine $A_4$

When projected onto the Coxeter plane, the 20 roots of $A_4$ reduce to 10 roots of $H_2$ because of the relations $k_1 - k_2 = \sigma(k_3 - k_5)$ +cyclic permutations. The roots of $H_2$ in Euclidean space $E_\parallel$ read

$$\pm\beta_1 = \pm\sqrt{2+\sigma}\,(k_5 - k_3) = \pm\sqrt{2+\tau}\,(k_1 - k_2)$$
$$\pm\beta_2 = \pm\sqrt{2+\sigma}\,(k_2 - k_5) = \pm\sqrt{2+\tau}\,(k_3 - k_4)$$
$$\pm(\tau\beta_1 + \beta_2) = \pm\sqrt{2+\sigma}\,(k_1 - k_3) = \pm\sqrt{2+\tau}\,(k_5 - k_4)$$
$$\pm(\beta_1 + \tau\beta_2) = \pm\sqrt{2+\sigma}\,(k_2 - k_4) = \pm\sqrt{2+\tau}\,(k_1 - k_5)$$
$$\pm\tau(\beta_1 + \beta_2) = \pm\sqrt{2+\sigma}\,(k_1 - k_4) = \pm\sqrt{2+\tau}\,(k_2 - k_3). \tag{10}$$

As we already noted in Section 2 the projected copies of the roots of $A_4$ are represented by the roots of $H_2$ either scaled by $\sqrt{2+\sigma}$ or $\sqrt{2+\tau}$ as can be seen from (10).

We first note from Fig. 4 that the line segments bisecting the roots $k_5 - k_4$ and $k_5 - k_2$ intersect at the point represented by the vector $k_5$ and the line segments bisecting the roots $k_2 - k_4$ and $k_3 - k_1$ intersect at the point represented by the vector $k_2 + k_3$. We will show that the affine reflections corresponding to these roots generate the representations of the group $H_2$ leaving the points $k_5$ and $k_2 + k_3$ invariant respectively.

We also note that the root $k_4 - k_1$ is orthogonal to the line segment joining the points $k_5$ and $k_2 + k_3$ which represents also the symmetry of the projected Voronoi cell. Let $(g, \alpha)$ be an element of the affine $H_2$ where $g$ is an element fixing the origin and the vector $\alpha$ representing a translation, is one of the projected root ($k_i - k_j$). The group element $(g, \alpha)$ can also be written in a matrix form. For example, the reflection with respect to the line segment bisecting the root $k_5 - k_4$ can be written as a $3 \times 3$ matrix (see Appendix A) :

$$C_1 = \frac{1}{2}\begin{bmatrix} -\sigma & -\sqrt{2+\tau} & -\sigma\sqrt{2} \\ -\sqrt{2+\tau} & \sigma & \sqrt{\frac{2}{2+\sigma}} \\ 0 & 0 & 2 \end{bmatrix}. \tag{11}$$

The $2 \times 2$ matrix $\frac{1}{2}\begin{bmatrix} -\sigma & -\sqrt{2+\tau} \\ -\sqrt{2+\tau} & \sigma \end{bmatrix}$ is the reflection with respect to the line segment orthogonal to the root $k_5 - k_4$ and passing through the origin. Therefore (11) represents a reflection with respect to the line segment through the origin and a translation by the vector $k_5 - k_4$. Any 2-component vector $(v_x, v_y)^T$ is replaced by $(v_x, v_y, 1)^T$ in this notation. As can be seen easily that the last column of the matrix is $\frac{1}{2}(-\sigma\sqrt{2}, \sqrt{\frac{2}{2+\sigma}}, 2)^T = k_5 - k_4$ representing a root vector. It is easy to show that (11) leaves the vector $k_5 = (\sqrt{\frac{2}{5}}, 0, 1)^T$ invariant. The group element representing the reflection with respect to the line segment passing through the origin, $k_5$ and $k_2 + k_3$ and which is also orthogonal to the root $k_4 - k_1$ can be written in this notation as

$$C_2 = \begin{bmatrix} 1 & 0 & 0 \\ 0 & -1 & 0 \\ 0 & 0 & 1 \end{bmatrix}. \tag{12}$$

Define the product $C_2 C_1 := (P, \lambda)$ where $P$ is a rotation around the origin and $\lambda = k_5 - k_1$ is a translation. The group element $(P, \lambda)$ can be written in the matrix notation as

$$(P, \lambda) = \frac{1}{2}\begin{bmatrix} -\sigma & -\sqrt{2+\tau} & -\sigma\sqrt{2} \\ \sqrt{2+\tau} & -\sigma & -\sqrt{\frac{2}{2+\sigma}} \\ 0 & 0 & 2 \end{bmatrix} = (P, k_5 - k_1), \tag{13}$$

where

$$P = \frac{1}{2}\begin{bmatrix} -\sigma & -\sqrt{2+\tau} \\ \sqrt{2+\tau} & -\sigma \end{bmatrix}, \tag{14}$$

represents the counterclockwise rotation by $\frac{2\pi}{5}$ around the origin. Then (13) represents the rotation by $\frac{2\pi}{5}$ around the point $k_5$. All group elements of $H_2$ fixing the point $k_5$ can be written as

$$I = \begin{bmatrix} 1 & 0 & 0 \\ 0 & 1 & 0 \\ 0 & 0 & 1 \end{bmatrix}, C_1 = \frac{1}{2}\begin{bmatrix} -\sigma & -\sqrt{2+\tau} & -\sigma\sqrt{2} \\ -\sqrt{2+\tau} & \sigma & \sqrt{\frac{2}{2+\sigma}} \\ 0 & 0 & 2 \end{bmatrix}, C_2 = \begin{bmatrix} 1 & 0 & 0 \\ 0 & -1 & 0 \\ 0 & 0 & 1 \end{bmatrix}, (P, k_5 - k_1),$$

$$(P^2, k_5 - k_2), (P^3, k_5 - k_3), (P^4, k_5 - k_4), C_2(P^2, k_5 - k_2), C_2(P^3, k_5 - k_3), C_2(P^4, k_5 - k_4). \quad (15)$$

The elements of $H_2$ fixing the point $k_2 + k_3$ can be generated by the elements $C_2$ and

$$C_3 = \frac{1}{2}\begin{bmatrix} -\sigma & \sqrt{2+\tau} & -\sqrt{2} \\ \sqrt{2+\tau} & \sigma & \frac{\tau\sqrt{2}}{\sqrt{2+\sigma}} \\ 0 & 0 & 2 \end{bmatrix}. \quad (16)$$

The matrix in (16) represents the reflection with respect to the line segment bisecting the root $k_2 - k_4$. The elements of the group $H_2$ leaving the point $k_2 + k_3$ invariant can then be written as

$$I, C_2, C_3, (P, (k_2 - k_4)), (P^2, \tau(k_2 - k_5)), (P^3, \tau(k_3 - k_5)), (P^4, (k_3 - k_1)),$$
$$C_2(P^2, \tau(k_2 - k_5)), C_2(P^3, \tau(k_3 - k_5)), C_2(P^4, (k_3 - k_1)). \quad (17)$$

The set of elements of (17) can also be obtained from the elements of (15) by a similarity transformation by translation element $t(\lambda)$ where $\lambda = k_2 + k_3 - k_5$. In (17) the group element $(P, (k_2 - k_4))$ represents a counterclockwise rotation by $\frac{2\pi}{5}$ around the point $k_2 + k_3$. When it is applied on Fig. 4 four times we obtain the tessellation given in Fig. 5, a five-fold symmetric patch with thin and thick rhombuses.

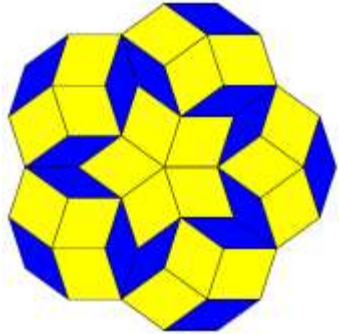

Figure 5. A patch obtained by rotating the patch in Figure 4 around the point $k_2 + k_3$.

Now, applying the reflection element $C_1$ of (15) or any other reflection element in (15) other than $C_2$ one enlarges the patch in Fig. 5. Further application of $(P, (k_2 - k_4))$ four times and subsequent application of $C_1$ and after several repetitions of the procedure one enlarges the patch in Fig. 5 and reaches something like the one in Fig. 6. This is a centrally symmetric patch consisting of thin and thick rhombuses fixing the point $k_2 + k_3$.

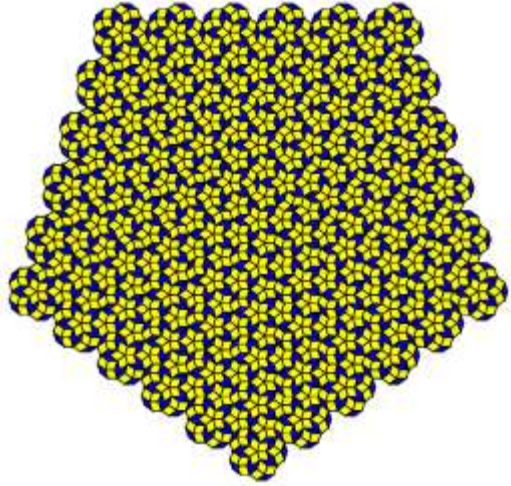

Figure 6. A centrally symmetric tessellation by thin and thick rhombuses around the point $k_2 + k_3$.

As we stated in Section 3 the Dynkin-diagram symmetry of $A_4$ induces the diagram symmetry of $H_2$ represented by $\eta$: $\beta_1 \leftrightarrow \beta_2$, $\gamma_1 \leftrightarrow \gamma_2$ which exchanges the vectors as $k_1 \leftrightarrow -k_4$, $k_2 \leftrightarrow -k_3$ and $k_5 \leftrightarrow -k_5$. Since $\eta^2 = 1$ this transformation extends the group $H_2$ to the group $H_2:Z_2$, a semi-direct product of $H_2$ with a group of order 2 denoted by $Z_2$ and generated by $\eta$. The new generator $\eta$ can be represented by the $3 \times 3$ matrix as

$$\eta = \begin{bmatrix} -1 & 0 & 0 \\ 0 & 1 & 0 \\ 0 & 0 & 1 \end{bmatrix}. \tag{18}$$

The group elements in (15) and (17) can be extended to a group of order 20 involving rotation elements of order 10.

A second set of Penrose tiling can be obtained by rotating the patch of Fig.5 four times by $\frac{4\pi}{5}$ by the elements in (15) leaving the point $k_5$ invariant. Then one can transform the original patch in Fig. 4 by elements of order 10 to fill the gaps and then applying successive rotations around the point $k_5$ leads to the finite fragment of the tessellation as shown in Fig. 7.

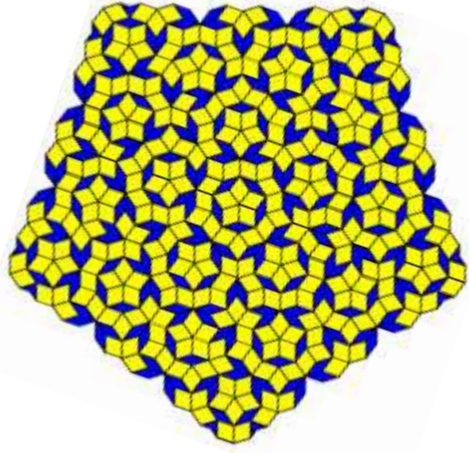

Figure 7. A centrally symmetric tessellation around the point $k_5$ by thick and thin rhombuses.

## 4. Tessellation by projection of the Voronoi cell $V(0)^*$ of the weight lattice $A_4^*$

The weight lattice $A_4^*$ is defined by the set of vectors $\frac{1}{5}\sum_{i=1}^{4} n_i \omega_i = \frac{1}{5}\sum_{i=1}^{5} m_i k_i$, $n_i, m_i \in \mathbb{Z}$. The vertices of the Voronoi cell $V(0)^*$ is the orbit of the vector [Koca et al, 2018]

$$\frac{1}{5}(\omega_1 + \omega_2 + \omega_3 + \omega_4) = \frac{1}{5}(4k_1 + 3k_2 + 2k_3 + k_4), \tag{19}$$

where $\frac{1}{5}(0 + \omega_1 + \omega_2 + \omega_3 + \omega_4)$ is the center of the fundamental simplex with vertices $(0, \omega_1, \omega_2, \omega_3, \omega_4)$. As such the fundamental simplexes are the Delone cells centralizing the vertices of the Voronoi cell $V(0)^*$. The affine $A_4$ operating on the fundamental simplex $(0, \omega_1, \omega_2, \omega_3, \omega_4)$ generate the weight lattice $A_4^*$. Noting that $k_1 + k_2 + k_3 + k_4 + k_5 = 0$ the vertex in (19) can be written as

$$\frac{1}{5}(5k_1 + 4k_2 + 3k_3 + 2k_4 + k_5). \tag{20}$$

The set of vertices of the Voronoi cell $V(0)^*$ is then the permutations of the vectors $k_i$ in (20). For further discussion, we shall drop the overall factor $\frac{1}{5}$. Then the permutations of the vectors $k_i$ in (20) is the realization of the permutohedron of degree 5 in 4D Euclidean space with the components given by $k_j = \sqrt{\frac{2}{5}}(e^{i\frac{2\pi}{5}j}, e^{i\frac{4\pi}{5}j})$. The permutohedron is abstractly defined as the permutation of the integers (54321) [Ziegler, G. M. 1995)] that is equivalent to the statement that the coefficients of the vectors $k_i$ are fixed but the vectors are permuted in (20). From now on we shall also drop the vectors $k_i$ and use just the coefficient of vectors in (20). Then the permutohedron defined as the permutations of the integers (54321) has $N_0 = 120$ vertices, $N_1 = 240$ edges, $N_2 = 150$ faces $= 60$(hexagons) $+ 90$ (squares), $N_3 = 30$ facets $= 10$(truncated octahedra) $+ 20$(hexagonal prisms). These numbers satisfy the Euler characteristic equation $N_0 - N_1 + N_2 - N_3 = 0$. Detailed structure of the permutohedron of degree 5 is interesting but we will not discuss here the details because we are interested in how the Voronoi cell $V(0)^*$ projects onto the Coxeter plane and how the tessellation is obtained by the action of the affine extension of $H_2$.

We note that projected images of the hexagonal and square faces come in two classes each. Let us consider the projections of the hexagonal faces. The following set of vertices obtained as the permutation of first three integers

$$(54321), (45321), (35421), (34521), (43521), (53421), \tag{21}$$

define a regular hexagon in 4D space. When projected onto the Coxeter plane it turns out to be a hexagon with edges $(\tau, \tau, 1, \tau, \tau, 1)$ (thick hexagon). A different hexagon is obtained when one considers the vertices of the hexagon given by

$$(52431), (42531), (32541), (32451), (42351), (52341). \tag{22}$$

This is also the permutations of the integers (543) but the integer 2 is at a different position. The projection of this hexagon will give us a hexagon whose edges can be represented by the numbers $(\tau, 1, 1, \tau, 1, 1)$ (thin hexagon). All 60 hexagons project onto one of these hexagons. The vertices representing squares project as follows. For example, the vertices (54321),(54231),(45231), (45321) forming a regular square in 4D projects onto a thin rhombus with edges (1,1,1,1). But the square represented by the vertices (53421), (43521), (42531) and (52431) projects onto a thick rhombus with edges $(\tau, \tau, \tau, \tau)$. So, 90 squares project onto one of these rhombuses or into a segment of the straight line. These hexagons and rhombuses are illustrated in Fig. 8. They are the only tiles obtained by projections of 150 faces. With these four different tiles we obtain the decagon in Fig. 8 as the projection of $V(0)^*$. It consists of 20 tiles, 5 tiles of each kind, and it is symmetric with respect to a line segment bisecting four different tiles.

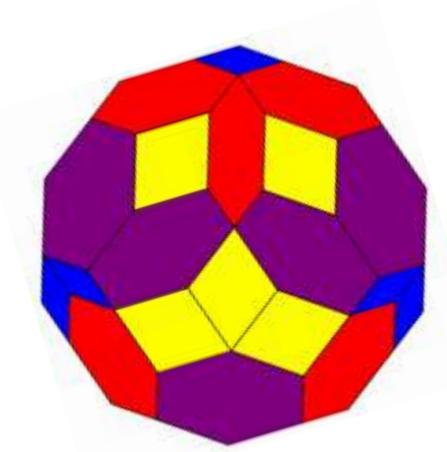

Figure 8. Decagon obtained as the projection of $V(0)^*$.

The group action similar to those discussed in Section 3 leads to the tessellation of the Coxeter plane by the four different tiles as shown in Fig. 9.

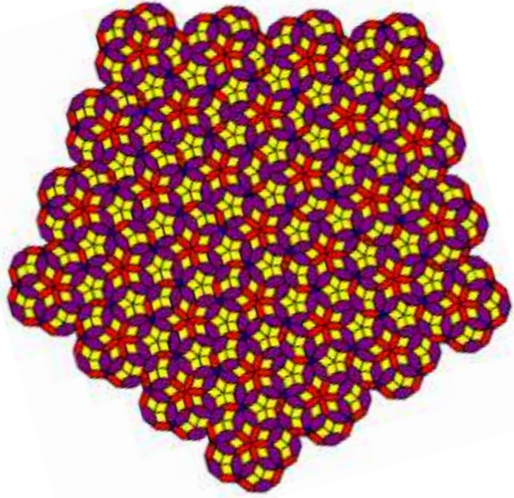

Figure 9. Tessellation of the plane by four tiles as the projection of the Voronoi tiling with $V(0)^*$.

## 5. Discussions

We have defined the affine extension of $H_2$ as the subgroup of affine $A_4$. The projection of the Voronoi cell $V(0)$ defines a decagon tiled by the projected rhombohedral facets in terms of thick and thin rhombuses as shown in Fig. 4. The same decagonal patch has been obtained earlier by different techniques [Gummelt, 1996; Baake et al, 1988]. Here we have studied it with an analysis of the group theoretical applications based on the affine $A_4$ and its affine subgroup $H_2$. We obtained two different centrally symmetric tilings by using two invariant points under the affine $H_2$. We have shown that any 5-fold symmetric point in the Coxeter plane can be determined as $\lambda = (n_1 + n_2)k_1 + n_2k_2 - n_3k_3 - (n_3 + n_4)k_4$. We have discussed the projection of the Voronoi cell $V(0)^*$ of the weight lattice leading to a decagonal patch consisting of hexagonal and rhombic tiles that has not been discussed elsewhere. Details of the mathematical work is presented in Appendix A.

## Appendix A. The affine $H_2$ as a subgroup of affine $A_4$:

Let $r_{\alpha,n}(\lambda) \coloneqq \lambda - ((\lambda, \alpha) - n)\alpha$, $n \in \mathbb{Z}$ represent the affine reflection with respect to the hyperplane $H_{\alpha,n} \coloneqq \{\lambda \in V | (\lambda, \alpha) = n\}$ where $\alpha$ is one of the roots of $A_4$. The generators $r_{\alpha_1,n_1}$, $r_{\alpha_2,n_2}$, $r_{\alpha_3,n_3}$ and $r_{\alpha_4,n_4}$ generate the affine $A_4$ whose point subgroup is isomorphic to the symmetric group $S_5$ fixing the point $\lambda$

$$\lambda = (n_1 + n_2)k_1 + n_2k_2 - n_3k_3 - (n_3 + n_4)k_4, \tag{A1}$$

and permuting the vertices of the Voronoi cell $V(\lambda)$. Here, the simple roots are given by $\alpha_1 = k_1 - k_2$, $\alpha_2 = k_2 - k_5$, $\alpha_3 = k_5 - k_3$, $\alpha_4 = k_3 - k_4$.

For $n_1 = n_2 = n_3 = n_4 = 0$, the generators represent the usual Coxeter-Weyl group leaving the origin invariant and permuting the vertices of the Voronoi cell $V(0)$.

Some useful relations are in order to clarify the further aspects of the affine $A_4$ and affine $H_2$:

$r_{\alpha_1,n_1} r_{\alpha_2,n_2}(\lambda) = r_{\alpha_2,n_2} r_{\alpha_1,n_1}(\lambda) = \lambda - (\lambda, \alpha_1)\alpha_1 - (\lambda, \alpha_2)\alpha_2 + n_1\alpha_1 + n_2\alpha_2$, for $(\alpha_1, \alpha_2) = 0$;

$r_{\alpha_2,n_2} r_{\alpha_1,n_1}(\lambda) = \lambda - (\lambda, \alpha_1)(\alpha_1 + \alpha_2) - (\lambda, \alpha_2)\alpha_2 + n_1(\alpha_1 + \alpha_2) + n_2\alpha_2$, for $(\alpha_1, \alpha_2) = -1$;

$r_{\alpha_1,n_1} r_{\alpha_2,n_2} r_{\alpha_1,n_1}(\lambda) = \lambda - (\lambda, (\alpha_1 + \alpha_2))(\alpha_1 + \alpha_2) + (n_1 + n_2)(\alpha_1 + \alpha_2) = r_{(\alpha_1+\alpha_2),n_1+n_2}$,

for $(\alpha_1, \alpha_2) = -1$. (A2)

Now we define the affine generators of $H_2$ as $R_1(n_1, n_3) := r_{\alpha_1,n_1} r_{\alpha_3,n_3}$ and $R_2(n_2, n_4) := r_{\alpha_2,n_2} r_{\alpha_4,n_4}$ satisfying the relations $R_1(n_1, n_3)^2 = R_2(n_2, n_4)^2 = 1$.

The Coxeter element of $A_4$ as well as of $H_2$ fixing the vector $\lambda$ in (A1) can be defined as

$R(n_1, n_2, n_3, n_4) := R_1(n_1, n_3) R_2(n_2, n_4)$ which transforms an arbitrary vector $\lambda$ as,

$R(n_1, n_2, n_3, n_4)(\lambda) = \lambda - (\lambda, (\alpha_1 + \alpha_2))\alpha_1 - (\lambda, \alpha_2)\alpha_2 - (\lambda, \alpha_2 + \alpha_3 + \alpha_4)\alpha_3 - (\lambda, \alpha_4)\alpha_4 + (n_1 + n_2)\alpha_1 + n_2\alpha_2 + (n_2 + n_3 + n_4)\alpha_3 + n_4\alpha_4$ with $R(n_1, n_2, n_3, n_4)^5 = 1$. (A3)

The Coxeter element represents a rotation of order 5 and for $n_1 = n_2 = n_3 = n_4 = 0$ it permutes the vectors $k_i (i = 1,2,3,4,5)$ in the cyclic order $R: k_1 \to k_2 \to k_3 \to k_4 \to k_5 \to k_1$. The reflection elements of $H_2$ with respect to certain line segments can be written as $R_1, R_2, R_1 R_2 R_1, R_2 R_1 R_2, R_1 R_2 R_1 R_2 R_1$ and $R^m, m = (1,2,3,4,5)$ are the rotation elements.

In terms of the roots of $A_4$ the reflection elements of $H_2$ are given by,

$$R_1(n_1, n_3) = r_{k_1-k_2,n_1} r_{k_5-k_3,n_3},$$
$$R_2(n_2, n_4) = r_{k_2-k_5,n_2} r_{k_3-k_4,n_4},$$
$$R_1 R_2 R_1 = r_{k_1-k_3,n_1+n_2+n_3} r_{k_5-k_4,n_3+n_4},$$
$$R_2 R_1 R_2 = r_{k_2-k_4,n_2+n_3+n_4} r_{k_1-k_5,n_1+n_2},$$
$$R_1 R_2 R_1 R_2 R_1 = r_{k_2-k_3,n_2+n_3} r_{k_1-k_4,n_1+n_2+n_3+n_4}.$$ (A4)

For $n_1 = n_3 = -1, n_2 = n_4 = 1, \lambda = k_2 + k_3$ the reflection elements in (A4) read

$R_1(1,1) = r_{k_2-k_1,1} r_{k_3-k_5,1}$: reflection with respect to the line segment from $k_2 + k_3 \to -k_5 - k_1$,

$R_2(1,1) = r_{k_2-k_5,1} r_{k_3-k_4,1}$: reflection with respect to the line segment from $k_2 + k_3 \to -k_5 - k_4$,

$R_1 R_2 R_1 = r_{k_3-k_1,1} r_{k_4-k_5}$: reflection with respect to the line segment from $k_2 + k_3 \to -k_1 - k_2$,

$R_2 R_1 R_2 = r_{k_2-k_4,1} r_{k_1-k_5}$: reflection with respect to the line segment from $k_2 + k_3 \to -k_3 - k_4$,

$R_1 R_2 R_1 R_2 R_1 = r_{k_1-k_4} r_{k_2-k_3}$: reflection with respect to the line segment from $k_2 + k_3 \to -k_2 - k_3$.

Any two reflection elements in (A5) generate the group $H_2$ fixing the vector $k_2 + k_3$. Similarly, the vector $k_5$ is left invariant by the group $H_2$ generated by $R_1(0,1) = r_{k_1-k_2}r_{k_5-k_3,1}$ and $R_2(1,0) = r_{k_5-k_2,1}r_{k_4-k_3}$.

**Translation element**:

The translation element of affine $A_4$ derived from Fig.1 reads $r_{k_1-k_4,1}r_{k_1-k_4}(\lambda) = \lambda + k_1 - k_4$ and the first two components $(k_1 - k_4)_{\|} = \frac{\tau(\beta_1+\beta_2)}{\sqrt{2+\sigma}}$ represents the translation element of the affine group $H_2$. For a general translation along $(k_1 - k_4)_{\|}$ take two reflection elements of $H_2$, $R(m_1, m_2) = r_{k_1-k_4,m_1}r_{k_2-k_3,m_2}$ and $R(0,0) = r_{k_1-k_4}r_{k_2-k_3}$. Then

$$R(m_1, m_2)R(0,0)(\lambda) = \lambda + m_1(k_1 - k_4) + m_2(k_2 - k_3)$$

represents the product of two translations in affine $A_4$. Since in the Coxeter plane $E_{\|}$, $(k_1 - k_4) = \tau(k_2 - k_3)$,

$$R(m_1, m_2)R(0,0)(\lambda) = \lambda + (m_1 + \tau m_2)(k_2 - k_3) = \lambda + (m_1 - \sigma m_2)(k_1 - k_4).$$

For $m_1 = 1, m_2 = 0$ one obtains the usual translation operator of $H_2$

$$R(1,0)R(0,0)(\lambda) = \lambda + (k_1 - k_4) = \frac{\tau(\beta_1+\beta_2)}{\sqrt{2+\sigma}}.$$

**Representations of the generators of affine $A_4$ by $5 \times 5$ matrices:**

A general affine transformation $(g, \lambda)$ of affine $A_4$ where $g$ is an element of $A_4$ fixing the origin and $\lambda$ represents the translation can be written in a $5 \times 5$ matrix form

$$\begin{bmatrix} g_{11} & g_{12} & g_{13} & g_{14} & \lambda_x \\ g_{21} & g_{22} & g_{23} & g_{24} & \lambda_y \\ g_{31} & g_{32} & g_{33} & g_{34} & \lambda_z \\ g_{41} & g_{42} & g_{43} & g_{44} & \lambda_w \\ 0 & 0 & 0 & 0 & 1 \end{bmatrix}$$

Let $a = \sqrt{2 + \tau}$. In this notation the generators of affine $A_4$ can be written as

$$r_{k_1-k_2} = \frac{1}{2a}\begin{bmatrix} a & \sigma & a & -\tau^2 & 0 \\ \sigma & \frac{\tau^4}{a} & -\sigma & -\frac{\tau}{a} & 0 \\ a & -\sigma & a & \tau^2 & 0 \\ -\tau^2 & -\frac{\tau}{a} & \tau^2 & \frac{\sigma^2}{a} & 0 \\ 0 & 0 & 0 & 0 & 2a \end{bmatrix}, \quad r_{k_2-k_5} = \frac{1}{2a}\begin{bmatrix} \sigma a & \tau & -a & -\tau^2 & 0 \\ \tau & \frac{2\tau+3}{a} & -\sigma & \frac{\tau}{a} & 0 \\ -a & -\sigma & a\tau & -1 & 0 \\ -\tau^2 & \frac{\tau}{a} & -1 & \frac{\tau+3}{a} & 0 \\ 0 & 0 & 0 & 0 & 2a \end{bmatrix},$$

$$r_{k_5-k_3} = \frac{1}{2a}\begin{bmatrix} \sigma a & -\tau & -a & \tau^2 & 0 \\ -\tau & \frac{2\tau+3}{a} & \sigma & \frac{\tau}{a} & 0 \\ -a & \sigma & a\tau & 1 & 0 \\ \tau^2 & \frac{\tau}{a} & 1 & \frac{\tau+3}{a} & 0 \\ 0 & 0 & 0 & 0 & 2a \end{bmatrix}, \quad r_{k_3-k_4} = \frac{1}{2a}\begin{bmatrix} a & -\sigma & a & \tau^2 & 0 \\ -\sigma & \frac{\tau^4}{a} & \sigma & -\frac{\tau}{a} & 0 \\ a & \sigma & a & -\tau^2 & 0 \\ \tau^2 & -\frac{\tau}{a} & -\tau^2 & \frac{\sigma^2}{a} & 0 \\ 0 & 0 & 0 & 0 & 2a \end{bmatrix},$$

$$r_{k_4-k_1,1} = \begin{bmatrix} 1 & 0 & 0 & 0 & 0 \\ 0 & -\frac{\tau}{a^2} & 0 & -\frac{2\tau}{a^2} & \frac{-\tau\sqrt{2}}{a} \\ 0 & 0 & 1 & 0 & 0 \\ 0 & -\frac{2\tau}{a^2} & 0 & \frac{\tau}{a^2} & -\frac{\sqrt{2}}{a} \\ 0 & 0 & 0 & 0 & 1 \end{bmatrix}. \tag{A6}$$

The automorphism group of the root system of $A_4$ is $A_4 : \mathbb{Z}_2$ and it is an extension by the Dynkin diagram symmetry given by

$$\delta = \begin{bmatrix} -1 & 0 & 0 & 0 & 0 \\ 0 & 1 & 0 & 0 & 0 \\ 0 & 0 & -1 & 0 & 0 \\ 0 & 0 & 0 & 1 & 0 \\ 0 & 0 & 0 & 0 & 1 \end{bmatrix}.$$

The generators of affine $H_2$ in affine $A_4$ are given by

$$R_1(0,0) = r_{k_1-k_2} r_{k_5-k_3} = \begin{bmatrix} -\frac{\tau}{2} & \frac{-\sqrt{2}+\sigma}{2} & 0 & 0 & 0 \\ \frac{-\sqrt{2}+\sigma}{2} & \frac{\tau}{2} & 0 & 0 & 0 \\ 0 & 0 & -\frac{\sigma}{2} & \frac{\sqrt{2}+\tau}{2} & 0 \\ 0 & 0 & \frac{\sqrt{2}+\tau}{2} & \frac{\sigma}{2} & 0 \\ 0 & 0 & 0 & 0 & 1 \end{bmatrix},$$

$$R_2(0,0) = r_{k_2-k_5} r_{k_3-k_4} = \begin{bmatrix} -\frac{\tau}{2} & \frac{\sqrt{2}+\sigma}{2} & 0 & 0 & 0 \\ \frac{\sqrt{2}+\sigma}{2} & \frac{\tau}{2} & 0 & 0 & 0 \\ 0 & 0 & -\frac{\sigma}{2} & -\frac{\sqrt{2}+\tau}{2} & 0 \\ 0 & 0 & -\frac{\sqrt{2}+\tau}{2} & \frac{\sigma}{2} & 0 \\ 0 & 0 & 0 & 0 & 1 \end{bmatrix},$$

$$R(1,0) = r_{k_4-k_1,1}r_{k_2-k_3} = \begin{bmatrix} 1 & 0 & 0 & 0 & 0 \\ 0 & -1 & 0 & 0 & \frac{-\tau\sqrt{2}}{a} \\ 0 & 0 & 1 & 0 & 0 \\ 0 & 0 & 0 & -1 & -\frac{\sqrt{2}}{a} \\ 0 & 0 & 0 & 0 & 1 \end{bmatrix}. \tag{A7}$$

$3 \times 3$ matrix representations of the generators of affine $H_2$:

$$R_1(0,0) = \begin{bmatrix} -\frac{\tau}{2} & \frac{-\sqrt{2+\sigma}}{2} & 0 \\ \frac{-\sqrt{2+\sigma}}{2} & \frac{\tau}{2} & 0 \\ 0 & 0 & 1 \end{bmatrix}, R_2(0,0) = \begin{bmatrix} -\frac{\tau}{2} & \frac{\sqrt{2+\sigma}}{2} & 0 \\ \frac{\sqrt{2+\sigma}}{2} & \frac{\tau}{2} & 0 \\ 0 & 0 & 1 \end{bmatrix}, R(1,0) = \begin{bmatrix} 1 & 0 & 0 \\ 0 & -1 & -\sqrt{\frac{2}{2+\sigma}} \\ 0 & 0 & 1 \end{bmatrix}. \tag{A8}$$

The Coxeter group $H_2$ can be extended to the automorphism group $H_2:\mathbb{Z}_2$ by the generator

$$\eta = \begin{bmatrix} -1 & 0 & 0 \\ 0 & 1 & 0 \\ 0 & 0 & 1 \end{bmatrix}.$$